\documentclass[conference]{IEEEtran}
\IEEEoverridecommandlockouts
\usepackage{cite}
\usepackage{amsmath,amssymb,amsfonts}
\usepackage{algorithmic}
\usepackage{graphicx}
\usepackage{textcomp}
\usepackage{xcolor}
\usepackage{cleveref}
\usepackage[utf8]{inputenc}
\usepackage{url}
\usepackage{caption}
\usepackage{subcaption}
\usepackage{ulem}
\usepackage{soul}
\usepackage{showexpl}

\usepackage[compact]{titlesec}  

\titlespacing{\section}{2pt}{1pt}{1pt}

\AtBeginDocument{
 \setlength\abovedisplayskip{2pt}
  \setlength\belowdisplayskip{0pt}
  }

\title{Normality of I-V Measurements Using ML 
\thanks{This research is sponsored by the INTERSECT Initiative as part of the Laboratory Directed Research and Development Program and by RAMSES project of Advanced Scientific Computing Research program, U.S. Department of Energy (DOE), and in part by the Office of Basic Energy Sciences, Division of Materials Sciences and Engineering, U.S. Department of Energy, and is performed at Oak Ridge National Laboratory managed by UT-Battelle, LLC, under contract DE-AC05-00OR22725 with the DOE. The US government retains and the publisher, by accepting the article for publication, acknowledges that the US government retains a nonexclusive, paid-up, irrevocable, worldwide license to publish or reproduce the published form of this manuscript, or allow others to do so, for US government purposes. 
}

}

\author{

\IEEEauthorblockN{
Anees Al-Najjar, Nageswara S. V. Rao, Craig A. Bridges, Sheng Dai}
\IEEEauthorblockA{\textit{Oak Ridge National Laboratory} Oak Ridge, TN, USA\\
\{alnajjaram,raons,bridgesca,dais\}@ornl.gov}
\and
\IEEEauthorblockN{
}
}

\begin{document}

\maketitle

\vspace*{-0.75in}

\section{Introduction}
\label{sec:introduction}
\begin{figure}[t]
\begin{center}
\vspace*{-.5in}
\includegraphics[width=0.4\textwidth]{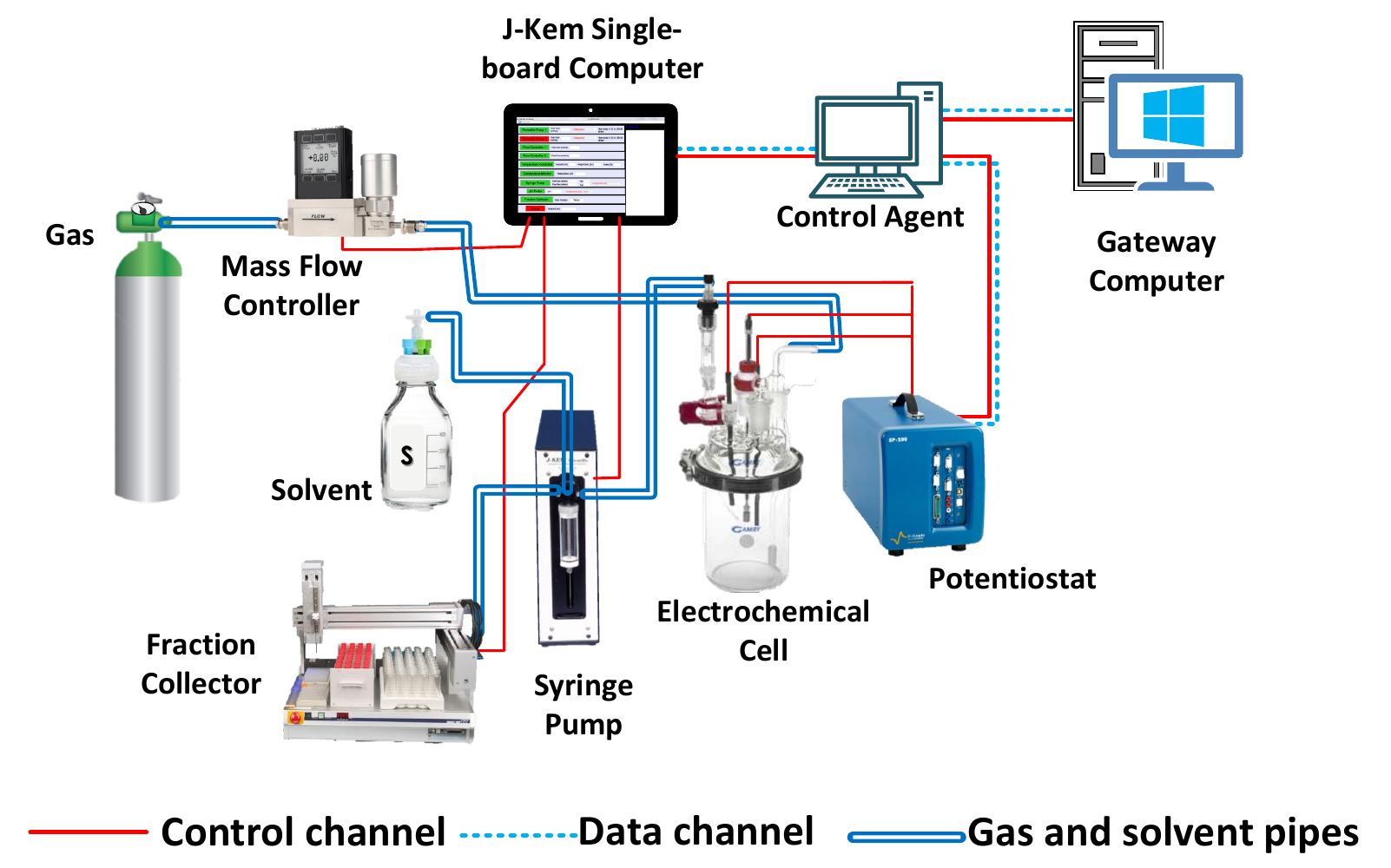}    
\end{center}
\caption{Electrochemistry workstation with potentiostat connected to an electrochemical cell and multiple instruments.} 
\label{fig:ACL_ElectroChemical_Setup}
\vspace*{-0.1in}
\end{figure}

There is an increased interest in instrument-computing ecosystems (ICEs) that support science workflows empowered by AI-automated experiments and computations in diverse areas.
In particular, electrochemistry ICEs are promising for accelerating the design and discovery of electrochemical systems for energy storage and conversion, by automating significant parts of workflows that combine synthesis and characterization experiments with computations.
They require the  integration of  flow controllers, solvent containers, pumps, fraction collectors, and potentiostats, all connected to an electrochemical cell, as illustrated in 
Fig.~\ref{fig:ACL_ElectroChemical_Setup}.
These are specialized instruments with custom software that is not originally designed for network integration.
We developed  network and software solutions for electrochemical workflows that adapt system and instrument settings in real-time for multiple rounds of experiments.
In particular, we developed Python wrappers for Application Programming Interfaces (APIs) of instrument commands and Pyro client-server modules that enable them to be executed from remote computers.
{The entire workflow is orchestrated by a Jupyter notebook running on a remote computer.}

We demonstrate this automated workflow by remotely operating the instruments and collecting their measurements to generate a voltammogram (I-V profile) of an electrolyte solution in an electrochemical cell.
These measurements are made available at the remote computing system and used for subsequent analysis. {\it It is important to ensure that measurements are collected under normal conditions throughout the automated workflow that may run unattended for days to weeks}.
In this paper, we focus on  a novel, analytically validated machine learning (ML) method for an electrochemistry ICE  to ensure that I-V measurements are consistent with the normal experimental conditions, and to detect abnormal conditions, such as disconnected electrodes or low cell content volume. 

\begin{figure}[b]
\centering
\vspace*{-0.1in}
\begin{subfigure}[c]{0.235\textwidth}
    \includegraphics[width=\textwidth,height=1.5in]{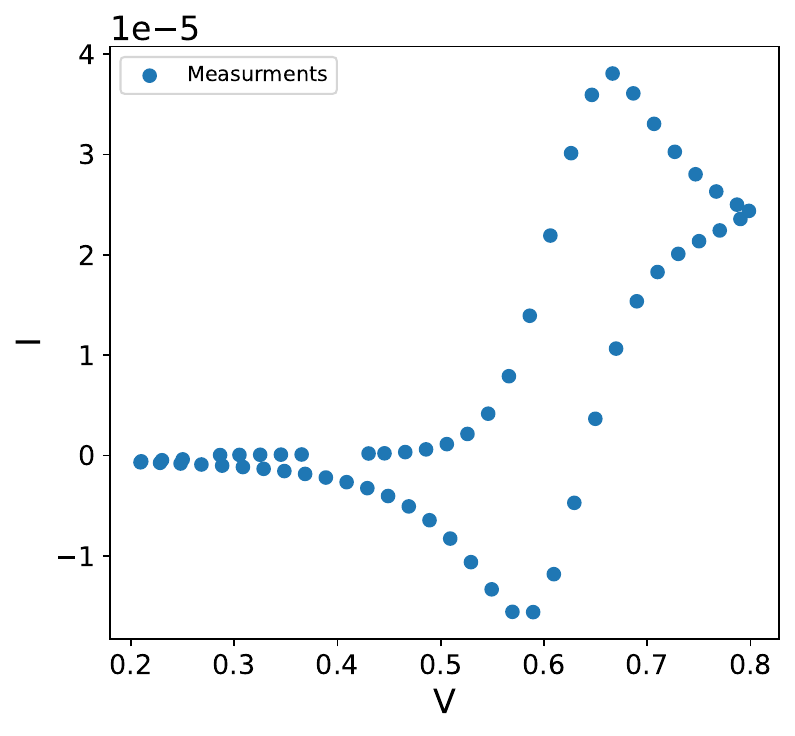}
\caption{continuous I-V profile}
\label{subfig:normal_iv_profile}
\end{subfigure}
\hfill
\begin{subfigure}[c]{0.23\textwidth}
         \includegraphics[width=\textwidth,height=1.5in]{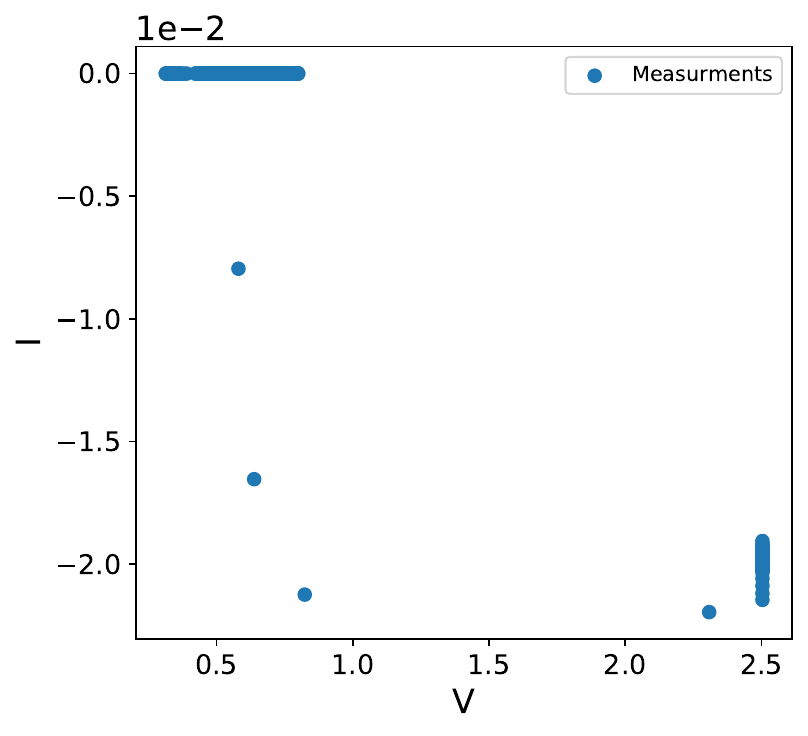}
          \caption{discontinuous I-V profile}
          \label{subfig:invalid_iv_profile}
\end{subfigure}
\hfill

\caption{I-V plot of measurements collected under normal and disconnected experimental  conditions.}
\label{fig:cv_feature_extraction}
\vspace*{-0.2in}
\end{figure}

\begin{figure*}[htb]
\centering
\vspace*{-0.2in}
\begin{subfigure}[c]{0.23\textwidth}
   \includegraphics[width=\textwidth,height=1.5in]{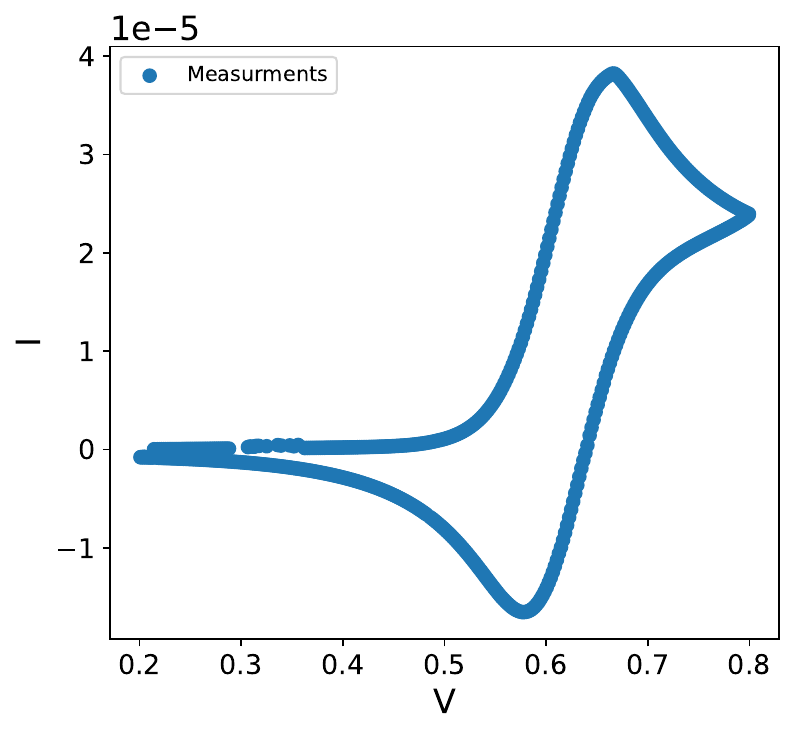}
\caption{{I-V measurements (N)}}
\label{subfig:testing_normal_iv_profile}
\end{subfigure}
\hfill
\begin{subfigure}[c]{0.23\textwidth}
        \includegraphics[width=\textwidth,height=1.5in]{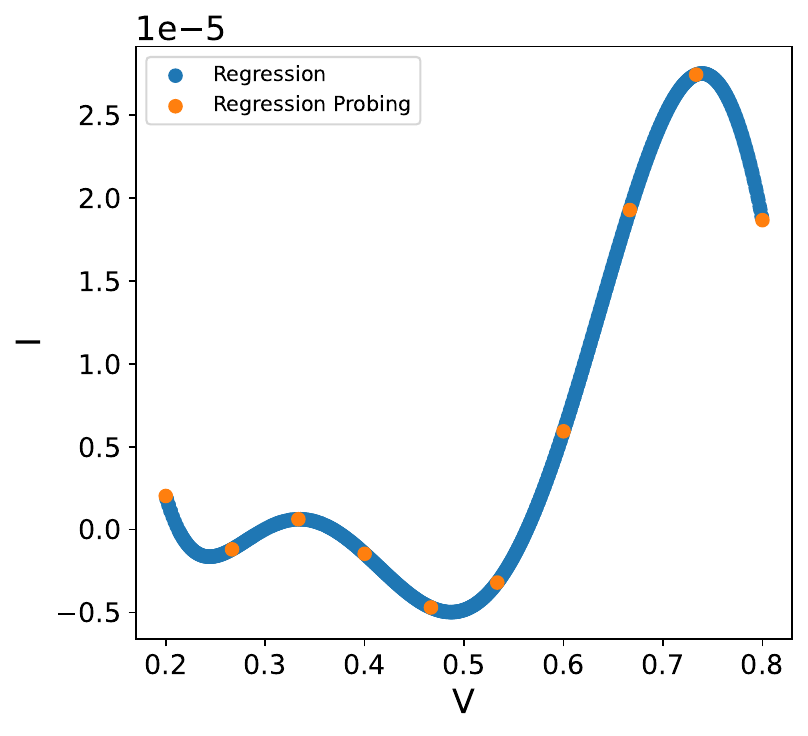}
\caption {{GPR fit, feature vector (N)}}
\label{subfig:classifying_testing_normal_iv_profile}
\end{subfigure}
\hfill
\begin{subfigure}[c]{0.23\textwidth}
   \includegraphics[width=\textwidth,height=1.5in]{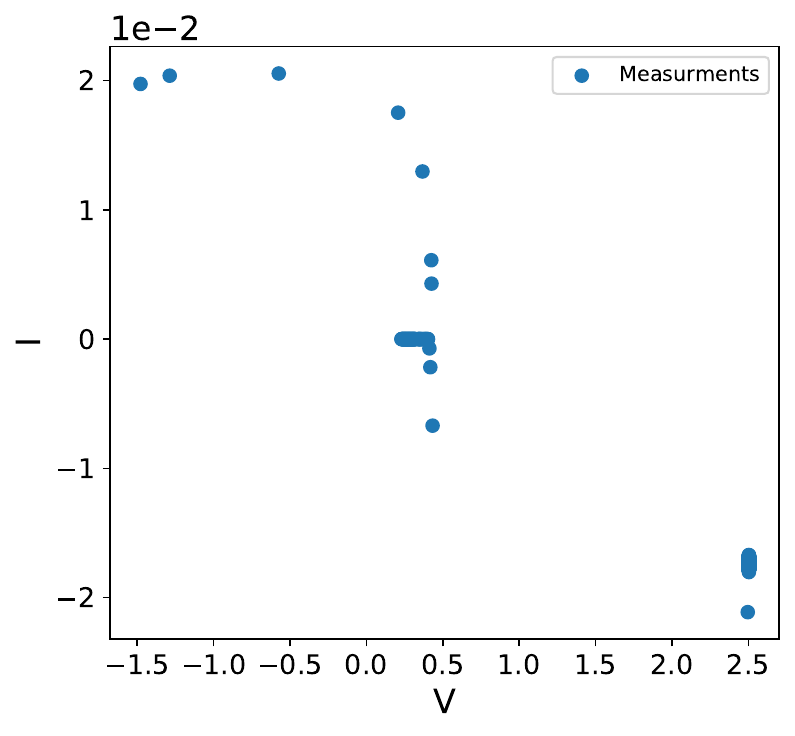}
\caption{{I-V measurements (D)}}
\label{subfig:testing_invalid_iv_profile}
\end{subfigure}
\hspace*{2em}
\begin{subfigure}[c]{0.23\textwidth}
        \includegraphics[width=\textwidth,height=1.5in]{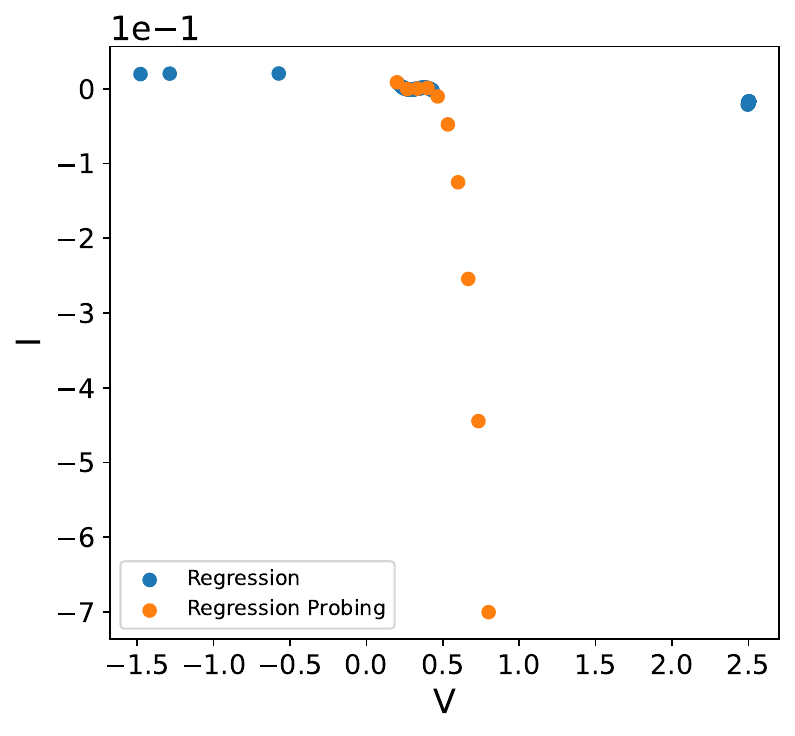}
\caption{{GPR fit, feature vector (D)}}
\label{subfig:classifying_testing_invalid_iv_profile}
\end{subfigure}
\hfill
\caption{{Normality testing of I-V measurements under normal (N) and disconnected (D) experimental conditions.}}
\label{fig:ml_results}
\vspace*{-0.2in}
\end{figure*}

\section{Electrochemistry  Ecosystem and Workflows}
\label{sec:ac_lab}

The  electrochemistry workstation incorporates a variety of instruments for liquid and gas transfers, along with a flexible electrochemical cell design to support testing, as illustrated in Fig.~\ref{fig:ACL_ElectroChemical_Setup}.
The cell is fed with gas (e.g., argon) from a gas tank via a computer-controlled mass flow controller, and is connected to pumps that dispense and withdraw liquids (e.g., filling the cell with solutions, washing the cell, or collecting fractions of liquid).    These instruments are directly connected to and controlled by a single-board computer controller to pump gas and solvent into the cell for conducting electrochemical tests of interest. The testing process itself takes place within the cell containing the electrolyte solution, which is initialized and controlled by a separate potentiostat instrument. Specifically, the potentiostat controls the electric potential (V) of a working electrode with respect to a reference electrode, and measures the current (I) flowing through the electrolyte solution; these measurements are used for generating its I-V profile.

The Cyclic Voltammetry (CV) is an electrochemistry technique in which potential is repeatedly applied to a working electrode immersed in an electrolyte solution, while simultaneously measuring the resulting current. Specifically, the potential is swept between two values at a controlled rate,  which results in an I-V cycle as shown in Figs.~\ref{subfig:normal_iv_profile} and~\ref{subfig:testing_normal_iv_profile}.
The I-V plot provides critical information about the electrolyte in the cell.  We exploit its geometric shape to develop a ML method to ensure that measurements reflect a normal experimental state, and detect abnormal conditions otherwise.

\section{Machine Learning for Normal I-V}
\label{sec:machine_learning}

Our custom ML method detects the deviations from normal I-V profiles due to conditions such as a disconnected electrode, low cell liquid level,  or other failures that produce ``abnormal'' voltammograms.
Examples of I-V measurements collected using the CV technique are shown in 
Fig.~\ref{subfig:normal_iv_profile} under a normal condition, where they form a (nearly) continuous closed curve.
When the electrode to the cell disconnected, their profile has a significantly different discontinuous shape as shown in Fig.~\ref{subfig:invalid_iv_profile}.
The ML method utilizes I-V measurements collected under both normal and disconnected conditions as a training set.


\subsection {Feature Design and Extraction: Classification}
\label{Feature_Design_and_Extraction}
A feature  vector is designed based on  I-V measurements by fitting  a continuous I-V regression curve with V and I as independent and dependent variables, respectively, and extracting 10 regression points at chosen V-values. This feature vector is
a 10-d vector with I-values computed at fixed probe points in V-space.
The Gaussian Process Regression (GPR) of data is used to obtain this feature vector of regression  I-values at chosen V values.
The GPR fits to I-V measurements and feature vectors at probe voltages under normal and disconnected conditions are shown in Fig.~\ref{subfig:classifying_testing_normal_iv_profile} and Fig.~\ref{subfig:classifying_testing_invalid_iv_profile}, respectively.

The feature vectors extracted under normal and disconnected conditions are used to train an Ensemble of Trees (EOT) classifier.
The output is 1 if the I-V measurements are consistent with normal measurements and 0 otherwise.

\subsection{Analytical Validation: Generalization Equations}
\label{Generalization_Equations}
This ML method is analytically validated by deriving the generalization equations
for GPR regression estimation and EOT classification, which have the same overall form \cite{algebra}.
In both cases, the size of data set used for estimation is $l$;
for GPR regression it corresponds to the number of individual I-V measurements used for feature vector estimation, and for EOT it corresponds to total number of normal and abnormal  I-V profiles used for training. 
The generalization equation  is of the form\\
\hspace*{0.7in}$
\mathbb{P}^l_{X,Y} \left [ I ( \hat{f}) - I( f^*) > \epsilon \right ] < \delta \left ( \epsilon, \hat{\epsilon}, l \right ),
$\\ 
where $\hat{f}$ is the GPR or EOT estimator, and $f^*$ is its best possible estimator that minimizes the expected error $I(.)$.
This equation guarantees that the expected error of estimator $\hat{f}$ is within that of optimal with confidence probability $\delta(.)$ which depends on $l$ and the training error $\hat{\epsilon}$ of $\hat{f}$.
The generalization equations have been very useful in establishing the solvability and assessing the performance of various ML methods\cite{algebra}.
Specifically, they guarantee this performance independent of and without the knowledge of the underlying data and error distributions.
For GPR, the confidence function is \\
\hspace*{0.7in}$ \delta_{GPR} = 8 \left ( \frac{32 \max(A,C)}{\epsilon} \right )^{2N_K} e^{- \epsilon^2l/512}, 
$\\
where $N_K$ is the number of component Gaussian functions and $A$ and $C$ are constraints that bound the estimators.
The confidence function of EOT classifier is given by\\
\hspace*{0.7in}$ \delta_{EOT}  = 8 g \left (  1+ \frac{ 256B N_L}{\epsilon} \right ) e^{-\epsilon^2 l/2048} ,$\\
where $B$ is a bound and $N_L$ is the number of EOT leaves.

\vspace*{0.1in}
\vspace*{-0.1in}
\section{Experimental Setup and Results}
\label{sec:Experimental_Setup}

The ICE implemented using API wrappers and Pyro modules automated the workflow for collecting I-V measurements, which are available at the remote computer via one-drive file system. The GPR method extracts the feature vector from them, which is used as input to the pre-trained EOT classifier that produces Boolean output.
Experimental results are shown in Fig.~\ref{fig:ml_results} for two typical cases.
Under normal experimental conditions (N) with all three electrodes in the cell connected properly to the potentiostat, I-V measurements and their GPR estimates form smooth curves, and the corresponding feature vector spans the GPR estimate as shown in Figs.~\ref{subfig:testing_normal_iv_profile} and \ref{subfig:classifying_testing_normal_iv_profile}; 
the EOT classifier output is 1 indicating the normality of this I-V profile.
Then, the reference electrode is disconnected (D) from the electrochemical cell and the I-V measurements are shown in Fig.~\ref{subfig:testing_invalid_iv_profile}, which result in a significantly different shape and values as compared to normal conditions.
Consequently, the corresponding estimated feature is significantly different as shown in Fig.~\ref{subfig:classifying_testing_invalid_iv_profile}, and  the classifier output is 0 indicating an abnormal condition.
This method is continuously used to check the normality of I-V measurements as they are  collected and transferred to remote computer throughout the workflow.


\section{Conclusions and Future work}
\label{sec:conclusions}

Our solution for developing ICE and supporting autonomous workflows is applicable to other science scenarios, including electron microscopy\cite{al2022enabling}. Integration of additional instruments and computing platforms, supporting more complex electrochemical workflows, and expandin


\bibliographystyle{ieeetr}
\bibliography{escienceposter23}

\end{document}